\documentclass[aps,reprint,amsmath, amssymb,groupedaddress,floatfix]{revtex4-1}
\usepackage{natbib}
\usepackage{graphicx}
\usepackage{bm}
\usepackage{color,soul}
\usepackage{hyperref}
\usepackage{float}
\usepackage{csquotes}
\usepackage{color,soul}
\usepackage{hyperref}
\hypersetup{colorlinks=true, urlcolor=blue, citecolor=blue}
\usepackage{braket}

\begin{document}

\preprint{APS}
\title{Spin Texture as Polarization Fingerprint of Halide Perovskites}
 \author{Mayank Gupta$^{1,2}$}
\author{B. R. K. Nanda$^{1,2,3}$}
\email{nandab@iitm.ac.in}
\affiliation{
  1. Condensed Matter Theory and Computational Lab, Department of Physics,
Indian Institute of Technology Madras, Chennai - 36, India\\
}
\affiliation{
2. Center for Atomistic Modelling and Materials Design,
Indian Institute of Technology Madras, Chennai - 36, India\\
}
\affiliation{
3. Functional Oxide Research Group,
Indian Institute of Technology Madras, Chennai - 36, India
}

\date{\today}

\begin{abstract}
Halide perovskites are perceived to be the promising class of materials for optoelectronics, spinorbitronics and topological electronics due to presence of strong spin-orbit coupling and polarization field. Here, we develop a Hamiltonian using quasi-degenerate perturbation theory to describe polarization field driven band structures and unique topological phases such as Dirac semimetallic rings for the universal class of Halide perovskites including both inorganic and hybrid members. The spin textures obtained through this theoretical model captures minute effects of polarization and hence can be used as a modern tool to determine the polarization field directions which have significant influence on spinorbitronics. The analysis brings out the concepts of hybrid spin textures intermixing the effects of valence and conduction bands under topological quantum phase transitions, and spin texture binaries where alignment of the spins are reversed between domains in the momentum space with sharp boundaries. The results are validated through density functional calculations.
\end{abstract}
\maketitle

\textbf{Introduction:} The physics of exploiting spin and orbital degrees of freedom of the electron has given rise to the emerging area of spinorbitronics. The center of it is the spin orbit coupling (SOC) which either occurs through the response of the electron spin to the nuclear electric field and/or the symmetry breaking polarization field. While the former has been the focus of examining topological electronic phases, the latter gives rise to Rashba, Dresselhaus or persistent type electric field tuned spin split \cite{Rashba,Dresselhaus,PST} and has gained attention after the giant Rashba spin splitting reported in ferroelectric GeTe \cite{GeTe1,GeTe2}. The electric field control spinorbitronics has practical advantage over traditional spintronics where the magnetization of spintronic compounds require large energies and might introduce undesired magnetic field \cite{stroppa2,bibes}.  The spin texture, which is intertwined with the polarization field and varies in the momentum space because of the same reason, has emerged as an excellent tool to demonstrate the complex interplay between spin, orbit, valley and lattice degrees of freedom in a crystal and in turn provide deep insight on the spinorbitronics and its tailoring. In optoelectronic and photovoltaic materials, the polarization field driven spin texture influences the recombination mechanism which in turn affects the carrier lifetimes \cite{lifetime1,lifetime2,lifetime3,lifetime4} and as a result the spin integrated optoelectronic is an emerging area of research.

Halide pervoskites (HP) lie at the crossroad of two of the phenomenal discoveries of the last two decades. These compounds - organic, inorganic or hybrid - are perceived as the next generation solar cell materials since the parity driven valence band  and conduction bands separated by a tunable bandgap with carrier lifetime give rise to excellent optoelectronic behavior \cite{opto1,opto2,opto3,opto4,opto5,opto6,opto7}. Since these systems has reasonable atomistic spin-orbit coupling, the pseudo-cubic symmetry of these systems can yield non-trivial topological insulating (TI) phase under pressure or strain \cite{RaviPRM,RaviPRB,RaviJCP}. 

When an organic molecule like CH$_3$NH$_3$ replaces an inorganic element like Cs at the A site of ABX$_3$, the inversion symmetry is broken to create a polarized electric field \cite{jinPNAS,KorePRB}. Also, there is a good number of reports suggesting the absence of a polarization field in such systems \cite{ferroelectric7,ferroelectric3,Frohna}. This may be due to macroscopic average led cancellation of the field whose direction randomly varies in each unit cell. However, the local symmetry breaking may give rise to local polarization fields, \cite{new_filippo} which can extend beyond the characteristic Rashba length scale (1-2 nm) \cite{lifetime1}. Furthermore, it has been suggested that, at low-temperature driven structural phase transition, the system may exhibit polarization \cite{new_filippo3,Stroppa}. Recent angle-resolved photoemission spectroscopy (ARPES) studies on CH$_3$NH$_3$PbBr$_3$ infers phase transition as well as Rashba splitting. In tetragonal CH$_3$NH$_3$PbBr$_3$, the Rashba type splitting of the top valence band in the momentum space is reported to be $k_0$ = 0.1 $\AA^{-1}$ and $E_R$ = 0.15 eV \cite{new_satoshi}.\par
Focusing on such intrinsic field and the future possibilities of tailoring the polarization, design principles have been developed for wider class of materials to create requisite field controlled and switchable spin texture \cite{zunger2020,LLTao}. However, presently the spin textures are either calculated from the eigenstates obtained using the simplistic \textbf{k$\cdot$p} theory or density functional calculations. While the former provides a very qualitative picture through approximated bands, the later, not being parametric, lack the ability to tune the spin texture for design purpose and fundamental understanding. Therefore, advanced parametric model Hamiltonians incorporating the polarization field and atomistic spin orbit coupling as well as the significant hopping interactions governing the band structure are very much necessary. Keeping this as the objective, in this work we have developed an elegant tight binding model Hamiltonian with the aid of quasi-degenerate perturbation theory which is applicable to demonstrate the variable spin texture of universal class of HP both in the normal and strain/pressure driven topological phases. The present study proposes that the spin texture can be decoded to get information related to basic structural symmetries, intrinsic polarization field and electronic structure of the perovskites. As a consequence, it
provides insights on the promising potentials of this family for applications in spinorbitronics, optoelectronics and topotronics. 

\begin{figure}
\centering
\includegraphics[angle=-0.0,origin=-1,height=6.5cm,width=7.5cm]{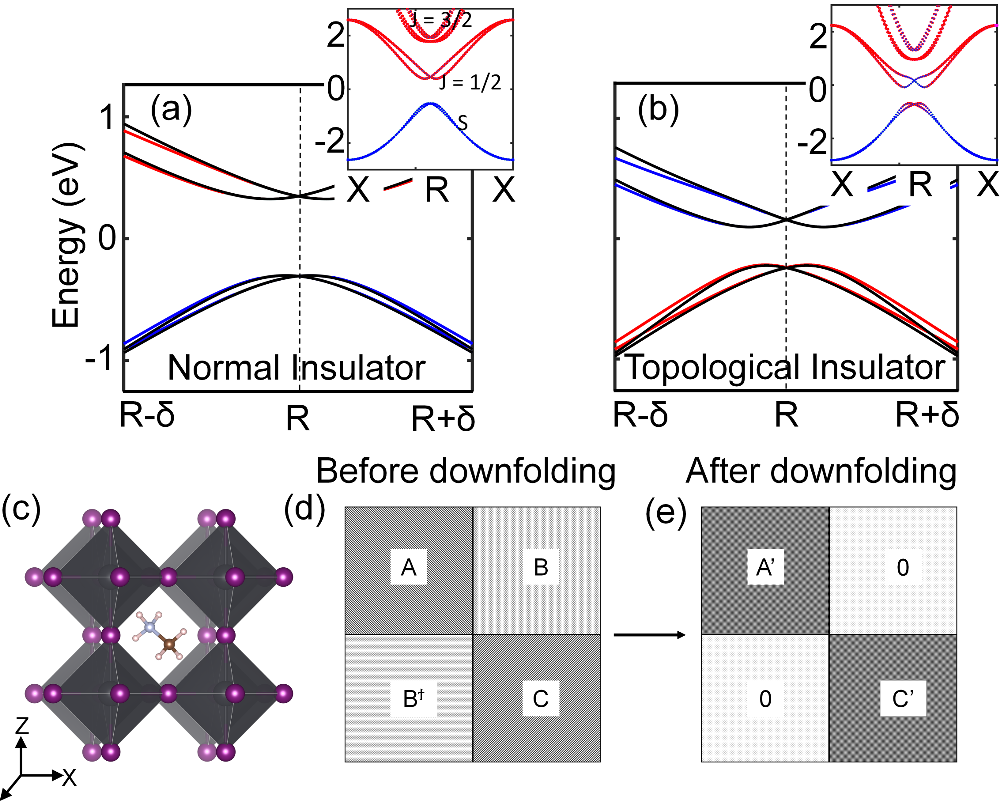}
\caption{Band structure of CH$_3$NH$_3$PbI$_3$ in the vicinity of Fermi level (a) in the ground state, and (b) under hydrostatic pressure. The coloured and black solid lines represent the band structure with and without downfolding of the Hamiltonian respectively. (c) The unitcell crystal structure of the compound. (d) Schematic representation of Lo\"wdin downfolding of the Hamiltonian (see Eq. 1) from a 8$\times$8 order to a 4$\times$4 order through the transformation as given in Eq. 6.}
\label{fig1}
\end{figure}

\textbf{Development of Model:} The prototypical band structure of non-centrosymmetric HP ABX$_3$ (e.g. CH$_3$NH$_3$PbI$_3$, Fig. \ref{fig1}(c)) is shown in Fig. \ref{fig1}(a). In the equilibrium state, the valence band edge is formed by the B-s states while predominantly the B-p states constitute the three conduction bands. The Rashba spin orbit coupling breaks the Kramer degeneracy around the time reversal invariant momenta (TRIM) R. Under hyrdrostatic pressure, the top valence band and the bottom conduction band invert their orbital characters significantly to induce the TI phase \cite{KorePRB}.
The model Hamiltonian, appropriate to describe the band structure for the family of HP, is formulated to be
\begin{equation}
      \begin{split}
      H = \sum_{i,a}\epsilon_{i a}c_{i a}^\dag c_{i a} + \sum_{ij;a b}t_{i a j b}(c_{i a}^\dag c_{j b}+h.c.) \\
       +\sum_{i,a b, \eta\eta'}\lambda_{a b \eta\eta'} (c_{i a\eta }^\dag c_{i b\eta'}+h.c.) + 
      \sum_{ij; a b}\alpha_{i a j b}^{P}(c_{i a}^\dag c_{j b}+h.c.).
      \end{split}
\end{equation}
Here, {\it i }({\it j}) and $a (b)$ are site and orbitals indices ($s$, $p_x$, $p_y$, and $p_z$) respectively corresponding to the B atom. The spin quantum number is represented by $\eta$ and $\eta'$ The parameters $\epsilon_{ia}$ and $t_{ia jb}$ respectively represent the onsite energy and hopping integrals and the latter are obtained through Slater-Koster formalism \cite{SLater_koster1}. The non-vanishing integrals in terms of the $\sigma$ and $\pi$ bonding strengths and direction cosines $l$, $m$, and $n$ are expressed as:
\begin{eqnarray}
    t_{s,s}&=&t_{ss},\\ \nonumber
    t_{s,p_x/p_y/p_z}&=&(l/m/n)t_{sp},\\\nonumber
    t_{p_x,p_x/p_y,p_y/p_z,p_z}&=&(l^2/m^2/n^2)t_{pp\sigma}\\\nonumber
    &&+(1-(l^2/m^2/n^2))t_{pp\pi}.\\ \nonumber
\end{eqnarray}

 Employing the tight-binding component of the Hamiltonian alone, the hopping parameters are estimated for a range of compounds \cite{RaviJCP} and results are listed in the supplementary information (SI) (See section- B, Table. S1 in SI), which we further averaged to develop an empirical relation among these  parameters as shown in Table- \ref{T1}.
\begin{table}
\centering
\caption {Average of TB interaction parameters in units of eV for the halide pervoskites ABX$_3$ family with cubic B lattice. In case of tetragonal and orthorhmobic B lattices, the $t$ can be made anisotropic.} 
\begin{tabular}{ccc}
\hline
Parameters &Average value& Scaled value\\
 \hline
$t_{ss}$& -0.2322 & 0.3$t$	\\
$t_{sp}$& 0.4833 & 0.6$t$ \\
$t_{pp\sigma}$ & 0.7944  &	$t$ \\
$t_{pp\pi}$ & 0.0989  &  0.13$t$	\\
\hline
\end{tabular}
\label{T1}
\end{table}
The parameter $t$ can be varied to represent the effect of external controls such as pressure. The onsite parameters $\epsilon_s$ and $\epsilon_p$, which represent the band center of the states after taking into account the B-${s,p}$ - X-$p$ interactions, are related to the bandgap ($E_g$) as \cite{RaviPRM} 
\begin{equation}
     \epsilon_p - \epsilon_s  = E_g - 6t_{ss} + 2t_{pp\sigma} + 4t_{pp\pi} + 2\lambda, 
\end{equation}  
Taking $\epsilon_s$ as reference (zero), $\epsilon_p$ can be related to the $E_g$.
The third and fourth terms of the Hamiltonian respectively represent the atomic spin-orbit coupling ( $\lambda \Vec{L} \cdot \Vec{S}$) and polarization field ($\vec{E}^P$) led spin-orbit coupling ($\alpha^P (\hat{E^P} \times\Vec{k})\cdot\Vec{\sigma}$). The value of $\lambda$ is taken to be 0.5eV which is of the same order as seen in MAPbI$_3$ \cite{KorePRB}. Here, $\alpha^P$ is the coupling strength which is proportional to the strength of the field, and $\sigma$ are the Pauli spin matrices.  The polarization field allows inter-orbital hopping \cite{Isabella} 
\begin{align}
    \alpha_{sp_x}^{P} &= (1-l^2)\gamma_{sp}^x - lm\gamma_{sp}^y - ln\gamma_{sp}^z \\
    \alpha_{p_xp_y}^{P} &= l\gamma_{pp}^y - m \gamma_{pp}^x  \nonumber
\end{align}
 The other elements such as $\alpha_{sp_y}$, $\alpha_{p_yp_z}$ can be written accordingly.  Here, $\vec{\gamma_{sp}}$ = ($\gamma_{sp}^x$, $\gamma_{sp}^y$, $\gamma_{sp}^z$) and $\vec{\gamma_{pp}}$ = ($\gamma_{pp}^x$, $\gamma_{pp}^y$, $\gamma_{pp}^z$) represent the hopping parameters due to an arbitrary field. For example, if the $\vec{E}^P$ is along the $z$-direction, then the component of the hopping vectors between two sites $\vec{R_i}$ (treated as center), $\vec{R_j}$ are expressed as 
\begin{eqnarray}
\gamma_{sp}^z &=& \langle s | E^zz | p_z, R_j^x\hat{x} \rangle = \langle s | E^zz | p_z, R_j^y\hat{y} \rangle \\ \nonumber
\gamma_{pp}^z &=& \langle p_x| E^zz | p_z, R_j^x\hat{x} \rangle = \langle p_y | E^zz | p_z, R_j^y\hat{y} \rangle
\end{eqnarray} 

The polarization field, which varies both in magnitude and direction due to random orientation of the organic molecule and hence random distortion of the BX$_6$ octahedra \cite{Frohna,Stroppa,Ong} influences the band structure significantly. Since, the direction of the field plays the most important role, for simplicity, we have taken $|\vec{\gamma_{sp}}|$ = $|\vec{\gamma_{pp}}|$ = $\gamma$. 
The complete Hamiltonian in the matrix form is provided in the SI (See section - A in SI). Altogether, $E_g$, $t$, $\lambda$ and $\gamma$, are the four parameters which can be tuned to explore emerging quantum phases through band structure and spin texture in the HP. As Eq. 5 suggests, the role of the polarization field is incorporated in the Hamiltonian through parameter $\gamma$. 

As seen from Fig. \ref{fig1}, $\ket{s_{\uparrow}}$ and $\ket{s_{\downarrow}}$ constitute the valence band edge (VBE), and $\ket{\frac{1}{2},\frac{1}{2}}$ and $\ket{\frac{1}{2},-\frac{1}{2}}$ constitute the conduction band edge (CBE). Since the other states hardly influence the topological phenomena, the spin texture, etc., the quasi degenerate perturbation theory \cite{Winkler} with the aid of Lo\"wdin downfold technique \cite{Lowdin,sashiPRL} can be applied to further reduce the effective Hamiltonian to a 4$\times$4 matrix consisting of both $S=1/2$ and $J=1/2$ manifold. Downfolding to $S=1/2$ or $J=1/2$ manifold separately fails to illustrate the topological phase of the system.  The downfolding process in general is demonstrated through Fig. \ref{fig1} (d, e) and Eq. 6. It says that if the Hamiltonian matrix H has blocks A, B, and C and we want to project the elements of B on A to create an effective block-diagonal matrix H$^{\prime}$, the elements of the block A$^{\prime}$ are given as
\begin{equation}
    A'_{ij} = A_{ij} + \sum_k \frac{B_{ik}B_{kj}}{\zeta - C_{kk}} 
    \approx A_{ij} + \sum_k \frac{B_{ik}B_{kj}}{\zeta_i - C_{kk}} 
\end{equation}
 In principle, A$^{\prime}_{ij}$ and $\zeta$ are obtained iteratively.  However, if $|B_{ik}| << |A_{ii} - C_{kk}|$, the iterative procedure is ignored. Here, $\zeta_i$ are the eigenvalue of full Hamiltonian and can be replaced by matrix diagonal element $A_{ii}$. We have numerically downfolded the Full Hamiltonian and the resulted band structure are shown in Fig. \ref{fig1} (a, b) which shows an excellent agreement around the Fermi level with the all electron band structure obtained from density functional theory (DFT) calculations.\\\\
 \begin{figure}
\centering
\includegraphics[angle=-0.0,origin=0.5,height=10.0cm,width=7.0cm]{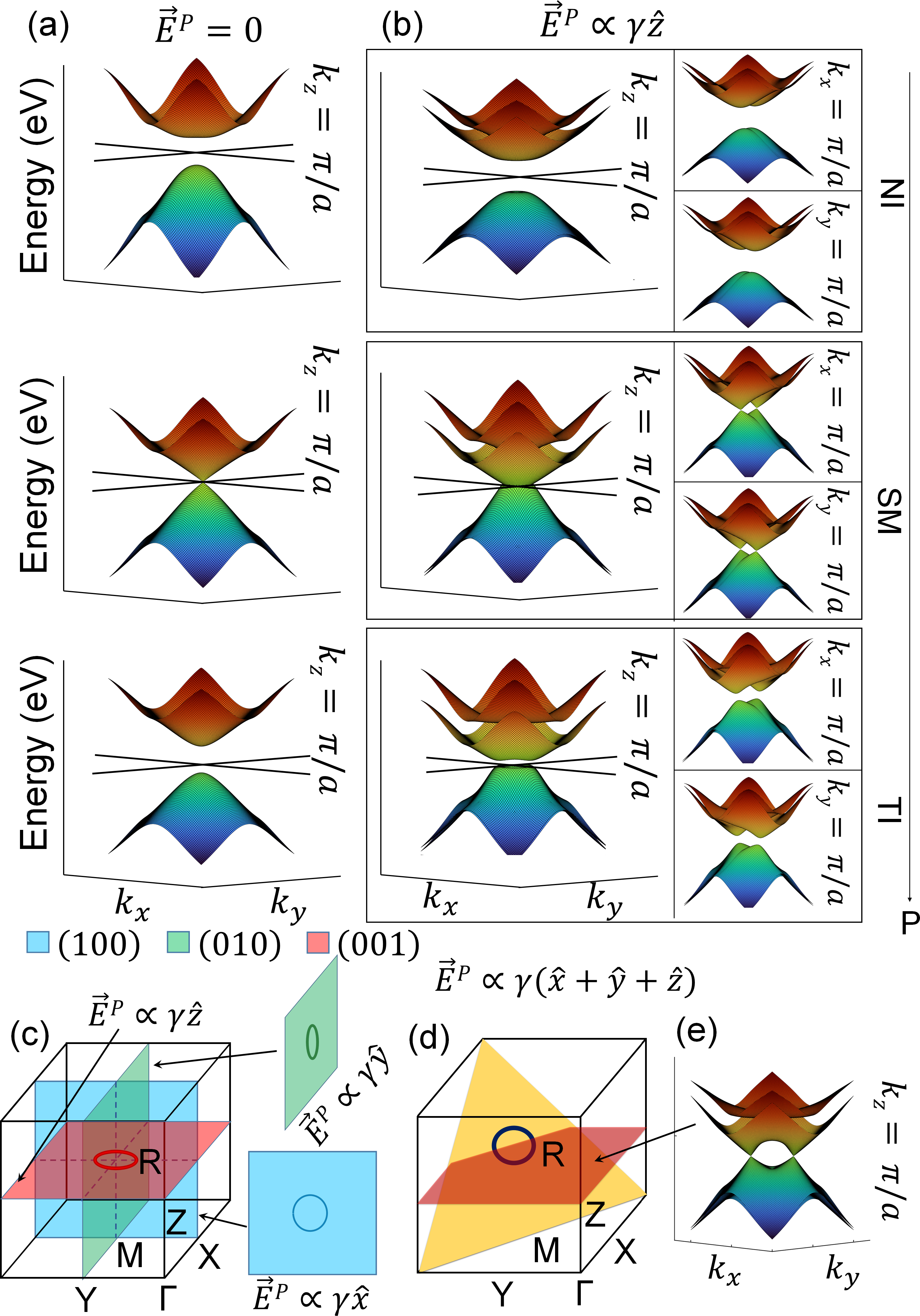}
\caption{The downfolded band structure of (a) centrosymmetric HP  and (b) non-centrosymmetric HP, with  along $\hat{z}$, plotted on the  $k_x-k_y$ plane and centered around the TRIM R ($\pi/a$, $\pi/a$, $\pi/a$). First row shows the ground state NI phase. At a critical pressure the HP exhibit the semimetallic phase (second row) and with further pressure, it stabilizes in the TI phase (third row). The inset shows the bands in the two other perpendicular planes. (c-d) The demonstration of Dirac semimetallic rings (DSR) around R and on the plane with normal along $\vec{E^P}$. (e) The band structure along the $k_x-k_y$ plane for a field $\vec{E^P}$ $\propto$ $\gamma(\hat{x}+\hat{y}+\hat{z})$.}
\label{fig2}
\end{figure}
\textbf{Computational details for DFT:}  The model study is further substantiated using density functional theory calculations. These are performed in pseudopotential-based projected augmented wave method as implemented on Vienna ab initio simulation package (VASP)\cite{Kresse1996}. Perdew-Burke-Ernzerhof generalized gradient approximation (PBE-GGA) \cite{Kresse1999,Blochl1994} is employed for the exchange interactions. The structural relaxation is carried out for each different orientation of the methylammonium (MA) molecule. For the atomic relaxation, $\Gamma$ centered k-point mesh of $8\times8\times8$ are used with energy cutoff of 450 eV in the absence of the spin-orbit coupling. Self-consistent field computations are performed for the band structure and spin texture calculations. For spin textures calculation, the k-point grid of total 441 k-points in three different xy, xz and yz planes around the TRIM R are designed. The spin textures are found by calculating expectation values ($\sigma_i, i = x, y, z$) of the Pauli spin matrix to the spinor eigenfunctions at every point of k-mesh grid. \\\\
\textbf{Results and analysis:} The downfolded Hamiltonian is further used to examine the role of $\vec{E^P}$ and pressure through Fig. \ref{fig2}. For convenience, first we shall consider $\vec{E^P}$ along $\hat{z}$-direction. In the normal insulator (NI) phase (upper panel) the $\vec{E^P}$  breaks the Kramer's degeneracy as expected. At a critical pressure, which is mapped to a set of critical hopping parameters, the semimetallic (SM) phase (middle panel) occurs. In the absence of the field, the valence band maximum (VBM) and conduction band minimum (CBM) touches each other only at the R point. With the field, the point translated to a ring on the $k_z$ = $\pi$/a plane with radius proportional to the field. Further increase in the pressure leads to band inversion giving rise to a TI Phase (lower panel). A detailed examination reveals that there always exist a critical pressure for any arbitrary $\vec{E^P}$ at which a SM phase is formed through Dirac semimetallic ring lying on a reciprocal plane whose normal is along the direction of the field. The demonstration of such rings for few selected $\vec{E^P}$ are shown in Fig. \ref{fig2} (c-d).

\begin{figure}
\centering
\includegraphics[angle=-0.0,origin=c,height=2.6cm,width=8.3cm]{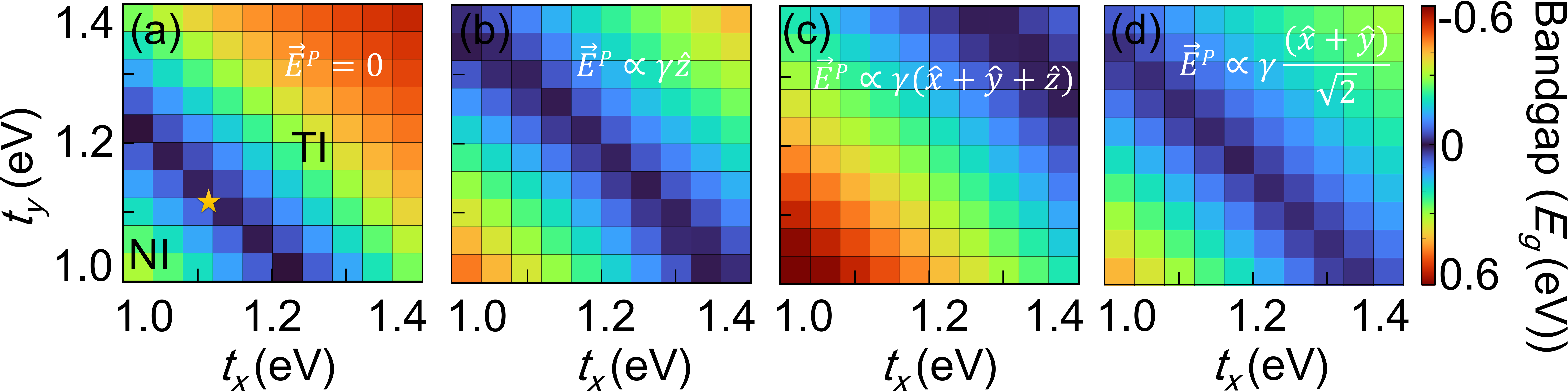}\\
\includegraphics[angle=-0.0,origin=c,height=3.5cm,width=8.0cm]{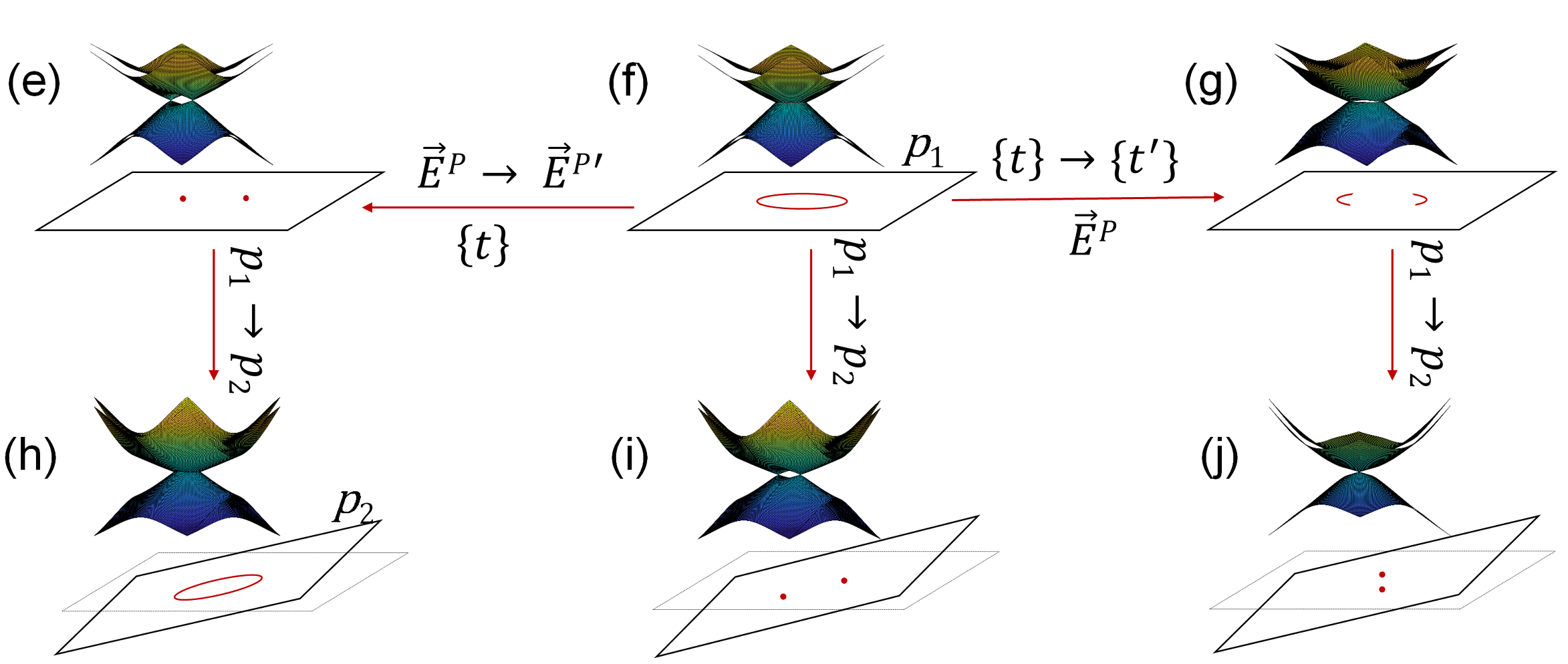}\\
\caption{(a-d) The mapping of the band gap in the $t_x$-$t_y$ plane while keeping the $t_z$ fixed at the critical value of 1.156 eV. The results are crucial for inferring the role of field and structural anisotropy on the electronic structure. (e-j) Schematic illustration of the perspective of the band structure with change in the field from $\vec{E^P}$ to $\vec{E^{P\prime}}$, and hopping interactions $\{t\}$ to $\{t^\prime\}$ and reciprocal plane $p_1$ to $p_2$.}
\label{fig3}
\end{figure}

Often structural anisotropies, which arise either due to natural growth process or through external effects such as strain, is accompanied with anisotropic hopping integrals that significantly influences the band structure. To examine if the SM phase survive due to such anisotropies, in Fig. \ref{fig3}, the bandgap is mapped in the $t_x$-$t_y$ plane keeping the $t_z$ fixed at the critical value $t_0$ (= 1.156 eV, when $\vec{E^P}$ = 0) for different field directions. Here, $t_x$ implies the hopping strength along the crystalline axis a$\hat{x}$. From Fig. \ref{fig3} the following salient features are obtained. (i) Irrespective of the field strength and direction, there exists a critical line in which the SM phase survives. The area under the line shows the NI phase and above it, the TI phase stabilizes. Identical observations are made in the $t_x$-$t_z$ and $t_y$-$t_z$ plane, keeping $t_y$ and $t_x$ fixed respectively. (ii) With increase in the field strength, the critical line shifts upward implying that one needs stronger hopping to achieve the NI to TI Phase transition. (iii) Since the SM phase is extremely sensitive to the hopping integral values, the gradual application of strain or pressure will not necessarily lead to a SM phase during NI-TI phase transition. Therefore, there will be systems in which the topological phase transition may occur through the first order phase transition.
In Fig. \ref{fig3}(e-j),  we schematically demonstrate the change in the perspective of the SM band structure with the change in the polarization field direction, hopping interactions and the reciprocal plane. For example, let us say that, when the polarization field is $\vec{E^P}$, a semimetallic ring is formed on the plane $p_1$ for a critical set of hopping interactions $\{t\}$. Now, if the field is changed to $\vec{E^{P\prime}}$, the Dirac circle is no longer observed on this plane and may occur some other plane $p_2$. Similarly, if $\{t\}$ changes to $\{t^{\prime}\}$ due to anisotropy, it may give rise to arcs on the same plane and Weyl points on some other plane $p_2$.
Overall, the field controlled bandgap in strained and/or pressured perovskites is turning out to be a robust tool to tune the optoelectronics and topological applications.

\begin{figure}
\centering
\includegraphics[angle=-0.0,origin=c,height=7.5cm,width=7.5cm]{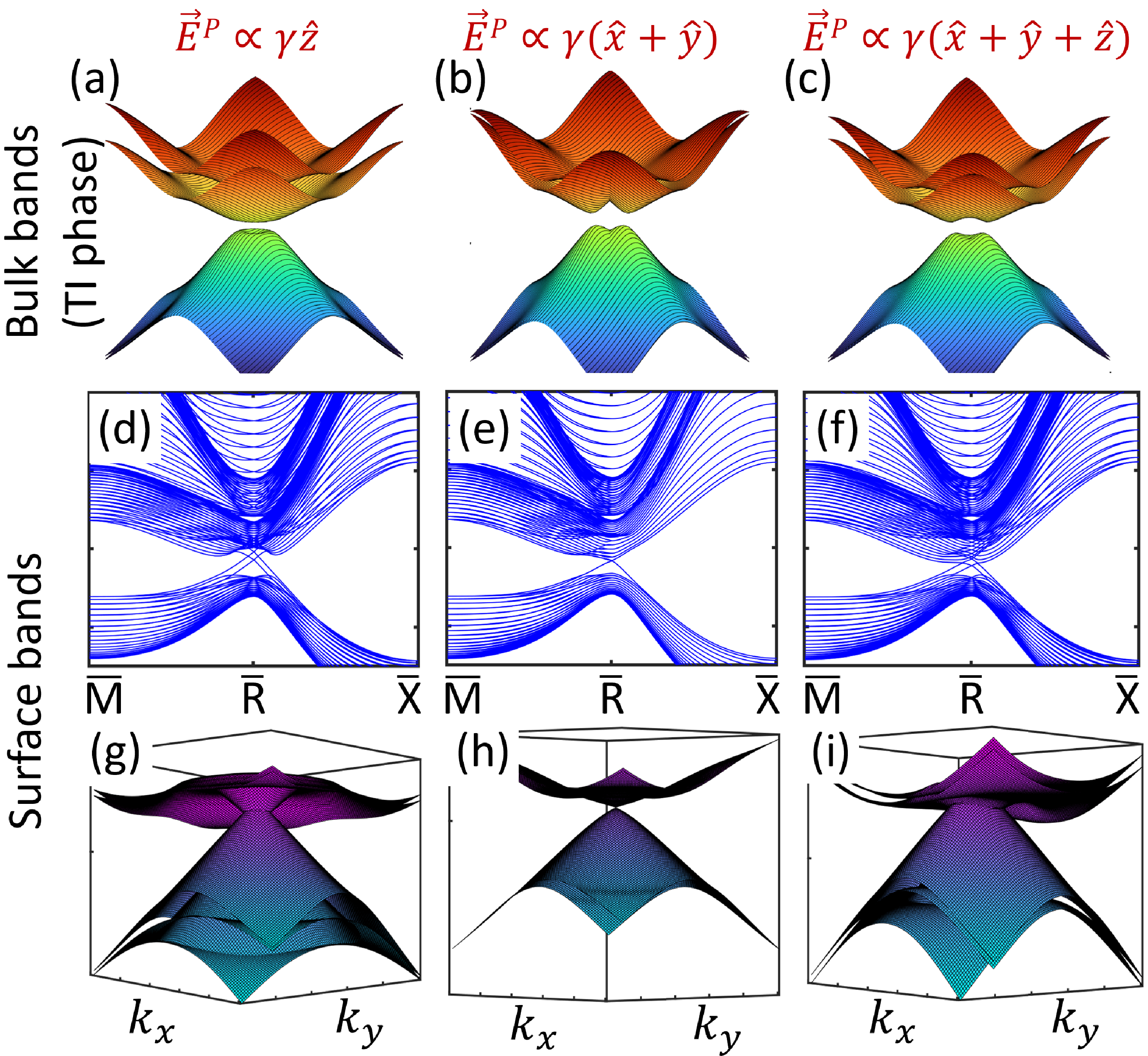}
\caption{(a-c) Bulk 3D bands in TI phase for three different polarization field. (d-f) Surface states obtained through the bulk-boundary correspondence, showing the various type of crossing surface states for different direction of polarization field. (g-i) 3D representation of surface states of corresponding field direction showing the different shape of band crossing.}
\label{fig4}
\end{figure}

In the TI phase unusual protected surface states are observed. By employing the TB model on a 25 layer slab grown along [001], we calculated the surface band structure for various field direction in the TI phase and results are shown in Fig. \ref{fig4}. For the field $\gamma\hat{z}$, two sets of Dirac cones interpenetrates to create a Dirac circle around the high symmetry point $\bar{R}$. Without the field, these two sets coincide to create a four-fold degenerate Dirac node \cite{RaviPRM,RaviPRB,KorePRB}. However, switching the direction of the field perpendicular to the growth direction, i.e. $\gamma(\hat{x}+\hat{y})$, then the formation of the Dirac circle is no longer observed.  Furthermore, we see that if the field  has a component along the growth direction, the penetration of the Dirac cones always occurs to form a Dirac curved circle. This implies that the shape of the protected surface  state is tunable to the polarization field.

\begin{figure}
\centering
\includegraphics[angle=-0.0,origin=-1.0,height=13.0cm,width=9cm]{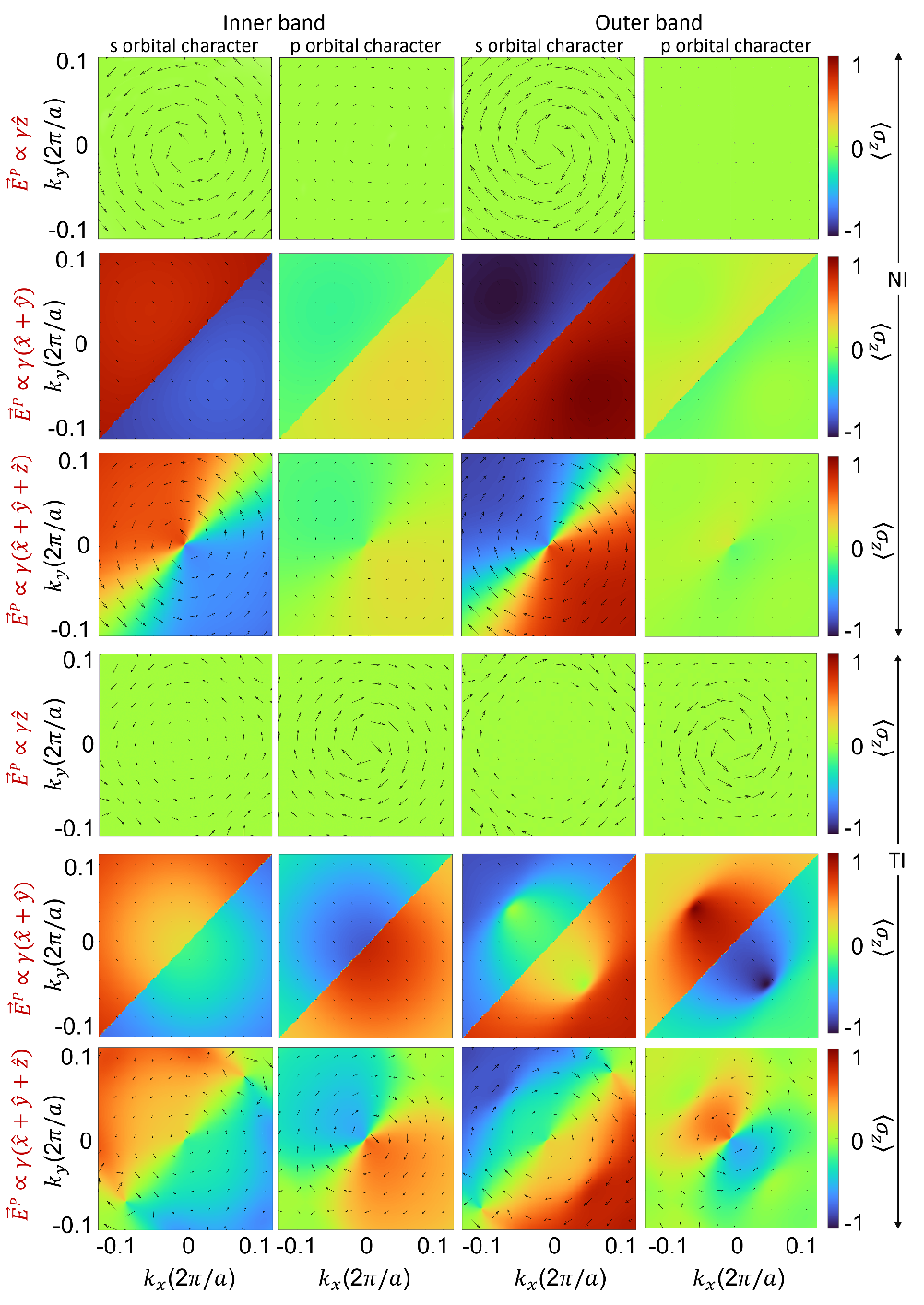}
\caption{The downfolding Hamiltonian obtained spin texture for the valence band near the R point for different direction of polarization field. (First row) Field in $\hat{z}$-direction, the spin texture is purely in-plane. (Second row) Field in [011] direction and (third row) field in [111] directions. The conduction band and valence bands spin texture are chiral to each other and hence only one of them is shown here. The spin texture is quite reversible.}
\label{fig5}
\end{figure}

As discussed, HP are the rare members which are capable of demonstrating both optoelectronic and topological phenomena. Therefore, the spin texture tuning is crucial in this class to realize unique spin-transport phenomena as well as efficient opto-electronic properties. Here, we evaluate the spin texture by taking the Pauli spin expectation of the downfolded bands under different field orientations. We find that the spin texture of the valence and conduction bands are chiral to each other and hence we will restrict our analysis to valence bands only and the results are shown in Fig. \ref{fig5}. When the field is along the $\hat{z}$-direction, in the NI phase, we observe the distinct Rashba spin texture both the inner and outer band (see first row) and are primarily formed by the $s$-orbital dominated eigen character. If the field is in the $xy$-plane, only the $\braket{\sigma_z}$ component survives and they are completely polarized with a diagonal boundary along the vector $\vec{k} = \vec{k_x} + \vec{k_y}$ (second row). When the field is mixed with components both along $z$-axis and in the plane, the Rashba helical spin texture become three dimensional (third row). The $\braket{\sigma_x}-\braket{\sigma_y}$ texture or $\braket{\sigma_z}$ texture is reversed by simply reversing the field along the $\hat{z}$-direction or in the $xy$-plane.

\begin{figure*}
\centering
\includegraphics[angle=-0.0,origin=-1.0,height=8.0cm,width=18cm]{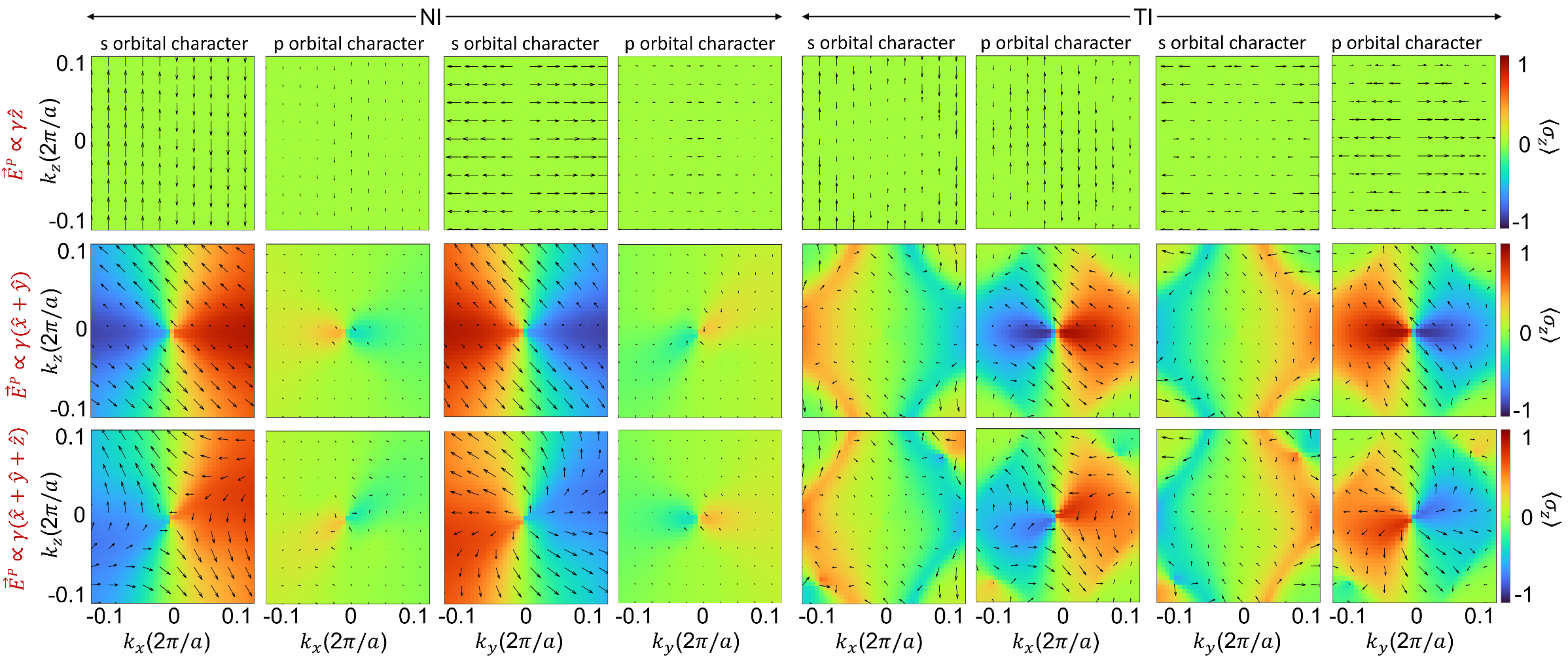}
\caption{The spin texture of outer valence band around the R-point in $k_x$-$k_z$ and $k_y$-$k_z$ plane for various polarization field direction. Field along [001]  (first row), (Second row) Field along [011] and (third row) field along [111]. The z component of spin expectation $\braket{\sigma_z}$ is shown by the background color.}
\label{fig6}
\end{figure*}
The topological insulating phase throws an interesting feature arising from the band inversion which makes the $J=1/2$ dominated chiral spin texture of the conduction bands penetrate the valence band spin texture breaking the Helical Rashba feature. If the field is in the $xy$-plane, the strength of $\braket{\sigma_z}$ is no longer uniform and modulated over the momentum space with multiple nodes. Such a hybrid spin texture is rarely observed.

The field dependent spin texture in the $k_x - k_z$ and $k_y - k_z$ plane for both NI and TI phase of outer valence band (Inner valence band; See SI, Fig. S2) are shown in Fig. \ref{fig6}. which are found to be much more unconventional and do not completely fit into earlier classifications of Rashba, Dresselhaus and persistent spin texture. However, since the valence \textcolor{blue}{and} conduction bands are parabolic around $R$, here we replace the orbital dependent Hamiltonian of Eq. 1 with  a simple \textbf{k$\cdot$p} Hamiltonian that can explain the cause of such textures.\cite{Vazna}
\begin{equation}
   H = \frac{\hbar^2k^2}{2m_e}I_{2\times2}+\alpha_{E^P}(\vec{E^P}\times\vec{k})\cdot\vec{\sigma}.\\
\end{equation}
Here, \textit{I} is the unit matrix and $\sigma = \{\sigma_x, \sigma_y, \sigma_z\}$ are the Pauli spin matrices. For perpendicular field, $\vec{E^P} = E_z\hat{z}$, the eigenvalues, eigenvectors and their Pauli spin expectations are
\begin{align}
    \varepsilon_R^{\perp}(k) = \frac{\hbar^2k^2}{2m_e} \pm \alpha_{E^P}\sqrt{k_x^2+k_y^2}\\ 
   \psi_R^{\perp}(k) = \mp\left(\begin{array}{c}
        \frac{k_y+ik_x}{\sqrt(k_x^2+k_y^2)} \\ \nonumber
        1 
   \end{array}\right), \\ \nonumber
   \braket{\sigma}_R^{\perp} \propto \pm \frac{2}{\sqrt{k_x^2+k_y^2}} \left(\begin{array}{c}
     -k_y \\
     k_x \\
     0
\end{array}\right)
\end{align}
Similarly, for the polarization field along the [110] direction, i.e. $\vec{E^P} = (E_x\hat{x}+E_y\hat{y}$), the eigenvalues, eigenvectors and their Pauli spin expectations are
\begin{eqnarray}
\varepsilon_{R}^{||}(k) &=& \frac{\hbar^2k^2}{2m_e}\pm \alpha_{E^P}\sqrt{(k_y-k_x)^2+2k_z^2}\\ \nonumber
\psi_R^{||}(k) &=& \left(\begin{array}{c}
     \frac{1+i}{2k_z}(k_y-k_x \pm\sqrt{(k_y-k_x)^2+2k_z^2})\\
     1 
\end{array}\right)\\ \nonumber
\braket{\sigma}_R^{||} &\propto& \pm \left(\begin{array}{c}
     \frac{(k_y-k_x)+\sqrt{(k_y-k_x)^2+2k_z^2}}{k_z} \\
     -\frac{(k_y-k_x)+\sqrt{(k_y-k_x)^2+2k_z^2}}{k_z}  \\
     \frac{(k_y-k_x)^2+(k_y-k_x)\sqrt{(k_y-k_x)^2+2k_z^2}}{k_z^2}
\end{array}\right)
\end{eqnarray}
The eigenvalues, eigenvectors at the $k_z$ = 0 can be calculated as,
\begin{gather}
\varepsilon_{R}^{||}(k_z = 0) = \frac{\hbar^2k^2}{2m_e}\pm \alpha_{E^P} (k_y-k_x)\\
\psi_R^{||},(k_z = 0) = \left(\left[\begin{array}{c}
     0\\
     1\\ 
\end{array}\right],\left[\begin{array}{c}
     1\\
     0\\ 
\end{array}\right]\right) \nonumber
\end{gather}
and thus the expectation values of Pauli spin matrices at the $k_z$ = 0 will be
\begin{equation}
\braket{\sigma}_R^{||}(k_z = 0) \propto \pm \left(\begin{array}{c}
     0 \\
     0 \\
     -1
\end{array}\right)
\end{equation}

For the perpendicular field alone, Eq. 8 describes the Helical nature of the $\braket{\sigma_x}$- $\braket{\sigma_y}$ spin texture in the $k_x-k_y$ plane (see Fig. \ref{fig5}, first row). This also tells that the texture is linear in the $k_x-k_z$ and $k_y-k_z$ planes as it is driven by $\braket{\sigma_y} = (\sigma_0)$ and $\braket{\sigma_x} = (-\sigma_0)$ respectively (see first row of Fig. \ref{fig6}, NI phase). When the field is planar, Eqs. 9 and 11 describes the spin texture. In the $k_x-k_y$ plane, only the $\braket{\sigma_z}$ survives which takes the sign $k_y-k_x$ and therefore forms a phase boundary at $k_x$ = $k_y$. (see second row of Fig. \ref{fig5}). If the field has both perpendicular and parallel components, then the resulted spin texture is almost a superimposition of the corresponding spin texture (see third row of Fig. \ref{fig5}). The spin textures obtained from the toy Hamiltonian are provided in SI (Fig. S1) and they have an excellent agreement with Fig. \ref{fig5}.

\begin{figure}
\centering
\includegraphics[angle=-0.0,origin=-1.0,height=12.5cm,width=8.0cm]{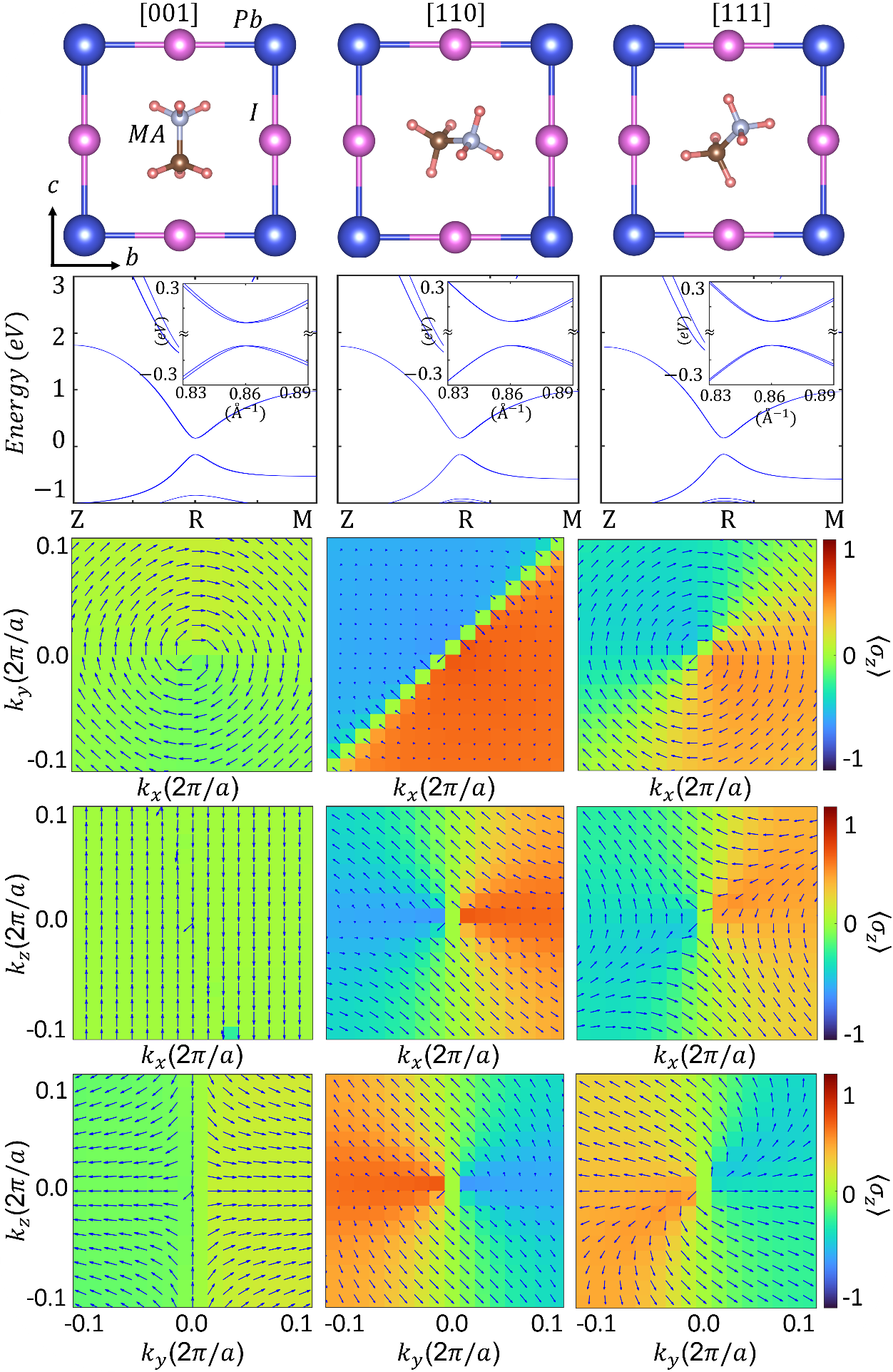}
\caption{ Spin texture obtained through DFT calculation for three different orientation of MA molecule inside the unit cell along with calculated band structure. Spin texture are plotted at three mutual perpendicular plane of Brillouin zone passing through the TRIM R. }
\label{fig7}
\end{figure}

To validate the model developed here and to emphasize the significance of spin texture in the HP family, in Fig. \ref{fig5}, \ref{fig6}, we present the DFT resulted band structure and spin texture for ideal cubic MAPbI$_3$ (a = 6.313 \AA) where the center of the unit cell coincides with the center of the MA molecule (CH$_3$NH$_3$). Such a system induces a very weak polarization field and that is directed along the C-N axis \cite{Stroppa}. The band structures corresponding to three MA orientations, along [001], [110], and [111], show that there is a very negligible Rashba splitting of the bands at R due to the weak polarization field which may not be observed through the band structure in the absence of highly dense $k$-mesh nearing the continuum.  However, the spin texture turned out to be very sensitive to the polarization field as can be clearly observed from Fig. \ref{fig7}. Furthermore, when compared with the spin texture of obtained from the downfolded model Hamiltonian for the polarization field direction along (001), (110), and (111) (see Fig. \ref{fig5}, \ref{fig6}), the DFT spin textures have an identical match which validates the model and make it a useful tool examine nature of polarization in the universal class of HP.

The significance of the spin texture is further realized when the ideal cubic structures of Fig. \ref{fig7} are relaxed to achieve the optimized orientations of MA molecule within the cubic environment. The DFT calculated spin texture are shown in the SI (Section - E, Fig. S3) and are replicated from the model Hamiltonian by tuning the field direction through $\gamma^i (i = x, y, z)$. This field direction agrees very well with that obtained through the polarization calculation using DFT. This suggests that the present model or its minor variations can be used to determine the polarization field direction through spin texture.

In conclusion, with an elegant minimal basis set based downfolded model Hamiltonian, taking into account the hopping interactions, atomistic and polarization field spin-orbit coupling,  we presented the band structure and spin texture for the universal class of HP. The downfolding is carried out using quasi degenerate perturbation theory. 
We demonstrate that the spin texture is ultra sensitive to determine the polarization field direction and independent of the field magnitude unlike the case of band structure where the magnitude plays the pivotal role. Hence, the spin texture captures very well the essence of polarization effect on the  the normal and topological insulating phase to serve as a finger print. The band inversion in the topological phase gives rise to unique hybrid spin texture where the presence of the $p$ characters in the valence band and the presence of $s$ character in the conduction band in the vicinity of TRIM R alters the spin texture observed in the NI Phase. Resembling the persistent spin texture, our calculations yield spin texture binaries where linear spins in one domain in the momentum space flips opposite after crossing a sharp boundary.   
The present studies also demonstrate the underlying mechanism and viability of a polarization field driven first order NI to TI quantum phase transition. Furthermore, polarization and pressure controlled distinct bulk and surface topological states such as Dirac semi-metallic closed and open contours and points are found to be tunable in this family. Existence of ferroelectric domains, crucial for achieving giant photovoltaic effect and high carrier lifetimes, in HP is a matter of debate \cite{RakitaPNAS,ferroelectric1,ferroelectric2,ferroelectric3,ferroelectric4,ferroelectric5,ferroelectric6,ferroelectric7,StrelcoveSA} and a widely pursued current research interest.

Efforts are also going on to tailor high efficiency perovskite solar cells through controlled doping \cite{XiaoNMAT} and by synthesizing pervoskites with varying organic cation frameworks \cite{ZhangJACS}. In recent times the emergence of 2D Halide perovskites stabilized through organic spacers will take the spin-orbit coupling physics to newer domains of fundamental and applied physics \cite{BlanconNN,MaoJACS}. Polarization is the common link to these above said research development and spin texture is going to become a bigger analytical and deterministic tool to tailor desired Halide perovskites. The spin texture is also highly significant towards exploring underlying physics of spinorbitronics and topotronics.

There are scopes to improve this model Hamiltonian presented here, so that the electronic structure and the polarization effect of the non pseudo-cubic Halide perovskites can be investigated. Also, for practical purposes, to address possible variations in the polarization field over space, one needs to incorporate the super cell approach and augment the Hamiltonian accordingly.\\

\textbf{Acknowledgements:} This work was funded by the Department of Science and Technology, India, through Grant No. CRG/2020/004330, and by Ministry of Education, India through Grant No. MoE/STARS-1/537. We also acknowledge the use of the computing resources at HPCE, IIT Madras.
%

\newpage
\widetext
\def\thesection{\Alph{section}}
\renewcommand{\thetable}{S\arabic{table}}
\renewcommand{\thefigure}{S\arabic{figure}}

\begin{center}
    \textbf{\Large Supplementary Information}
\end{center}

\section{Tight Binding Hamiltonian}
The complete tight binding Hamiltonian as explained in main text is, 
\begin{equation}
      \begin{split}
      H = \sum_{i,a}\epsilon_{i a}c_{i a}^\dag c_{i a} + \sum_{ij;a b}t_{i a j b}(c_{i a}^\dag c_{j b}+h.c.) \\
       +\sum_{i,a b, \sigma\sigma'}\lambda_{a b \sigma\sigma'} (c_{i a\sigma }^\dag c_{i b\sigma'}+h.c.) + 
      \sum_{ij; a b}\alpha_{i a j b}^{P}(c_{i a}^\dag c_{j b}+h.c.).\\
      \end{split}
\end{equation}
The components of this Hamiltonian can be understand as
\begin{equation}
    H = H_{TB} + H_{SOC} + H_{E^P} \\ 
\end{equation}
In matrix form, each term of Hamiltonian are given as, 
\[ 
\mathbf{H_{TB}+H_{SOC}}
 =  \left( \begin{array}{cc}
    \mathbf{H_{\uparrow\uparrow}}& \mathbf{H_{\uparrow\downarrow}} \\
    \mathbf{H_{\downarrow\uparrow}^{\dagger}}&  \mathbf{H_{\downarrow\downarrow}}
\end{array}  \right), 
\]
\[
\mathbf{ H_{\uparrow\downarrow}}= (\mathbf{H_{\downarrow\uparrow}})^\dagger
 =  \left( \begin{array}{cccc}
    0 & 0 & 0 & 0\\
    0 & 0 & 0 & \lambda\\ 
    0 & 0 & 0 & -i\lambda\\
    0 & \lambda & -i\lambda & 0\\ 
\end{array}  \right)
\]

\[
\mathbf{ H_{\uparrow\uparrow}}= \mathbf{(H_{\downarrow\downarrow})^{*}}
 =  \left( \begin{array}{cccc}
    \epsilon_s+f_0 & 2it_{sp}S_x & 2it_{sp}S_y & 2it_{sp}S_z\\
    -2it_{sp}S_x & \epsilon_p+f_1 & -i\lambda &0\\ 
    -2it_{sp}S_y & i\lambda & \epsilon_p+f_2 & 0\\
    -2it_{sp}S_z &   0 & 0 & \epsilon_p+f_3 
\end{array}  \right)
\]
Here, $S_{x/y/z}$ = sin(x/y/z), $C_{x/y/z}$ = cos(x/y/z) and  $\epsilon_s$ and $\epsilon_p$ are effective on-site energies and $t_{ss}$, $t_{sp}$, $t_{pp\sigma}$ and $t_{pp\pi}$ are hopping interactions between different orbitals and, 
\begin{eqnarray}
f_0& = &2t_{ss}(C_x+C_y+C_z) \nonumber\\
f_1& = &2t_{pp\sigma}C_x+2t_{pp\pi}(C_y+C_z) \nonumber \\
f_2& = &2t_{pp\sigma}C_y+2t_{pp\pi}(C_x+C_z) \nonumber\\
f_3& = &2t_{pp\sigma}C_z+2t_{pp\pi}(C_x+C_y).
\end{eqnarray}

The dispersion relations are obtained from the Slate-Koster tight binding formalism \cite{SLater_koster1}. Also, the Hamiltonian matrix for polarized field for each direction of polarization. The tight binding Hamiltonian for the field in each $\hat{x}$, $\hat{y}$ and $\hat{z}$ direction can be written as \cite{Isabella}
\begin{equation}
 H^{x}
 =  \sum_{\langle j \rangle}e^{i\Vec{k}\Vec{R_j}} \left( \begin{array}{cccc}  
 
    0 & (1-l^2)\gamma^x_{sp} & -lm\gamma^x_{sp} & -ln\gamma^x_{sp}\\
    (1-l^2)\gamma^z_{sp} & 0 & -m\gamma^x_{pp} & -n\gamma^x_{pp}\\ 
    -lm\gamma^x_{sp} & m\gamma^x_{pp} & 0 & 0\\
     -lm\gamma^x_{sp} &  n\gamma^x_{pp}  & 0 &0 \\
\end{array}  \right),
\end{equation}

\begin{equation}
 H^{y}
 =  \sum_{\langle j \rangle}e^{i\Vec{k}\Vec{R_j}} \left( \begin{array}{cccc}  
 
    0 & -lm\gamma^y_{sp} & (1-m^2)\gamma^y_{sp} & -mn\gamma^y_{sp}\\
    -lm\gamma^y_{sp} & 0 & l\gamma^y_{pp} & 0\\ 
    (1-m^2)\gamma^y_{sp} & -l\gamma^y_{pp} & 0 & -n\gamma^y_{pp}\\
     -mn\gamma^y_{sp} &  0  & n\gamma^y_{pp} &0 \\
\end{array}  \right),
\end{equation}

\begin{equation}
\small
 H^{z}
 = \sum_{\langle j \rangle}e^{i\Vec{k}\Vec{R_j}} \left( \begin{array}{cccc}  
 
    0 & -ln\gamma^z_{sp} & -mn\gamma^z_{sp} & (1-n^2)\gamma^z_{sp}\\
    -ln\gamma^z_{sp} & 0 & 0 & l\gamma^z_{pp}\\ 
    -mn\gamma^z_{sp} & 0 & 0 & m\gamma^z_{pp}\\
     (1-n^2)\gamma^z_{sp} &  -l\gamma^z_{pp}  & -m\gamma^z_{pp} &0 \\
\end{array}  \right),
\end{equation}

The resultant Hamiltonian for polarized field with field in random direction can be simplify as given in following expression,
\begin{equation}
    H_{E^P} = \gamma_x H^{x} + \gamma_y H^{y} + \gamma_z H^{z},
\end{equation}
Here, $\gamma_x, \gamma_y $ and $ \gamma_z$ are the coupling strength due to the polarization in [100], [010] and [001] direction respectively. Furthermore, for the simplification of our model we have considered the $\gamma_{sp}^i = \gamma_{pp}^i$ (i = x, y, z). When the results of the model are compared with that of the DFT, the $\gamma$s are optimized to replicate the DFT obtained polarized field (it is discussed in detail in section - E). For regular analysis, $\gamma^i$ are treated as a single parameter $\gamma$.
\section{ Estimated tight binding Parameters}
In the main text, an empirical relation between the tight binding interaction parameters is presented. This relation was established by fitting of the TB band structure with that of the DFT for a set of cubic halide pervoskites \cite{RaviJCP}. The resulted TB parameters are given in Table. \ref{T1}.
\begin{table}[ht]
\centering
\caption{ Interaction parameters ($\epsilon$s and $t$s) and SOC strength $\lambda$ for cubic phases in units of eV.}
\begin{tabular}{cccccccccc}
\hline
\hline
B&X & $\epsilon_s$ & $\epsilon_p$ & $t_{ss}$ & $t_{sp}$& $t_{pp\sigma}$ & $t_{pp\pi}$ & $\lambda$  \\
 \hline
&Cl&1.17 &	6.47&	-0.26&	0.47&	0.75&	0.09&	0.07\\
Ge&Br&1.46&	6.10&	-0.23&	0.48&	0.84&	0.09&	0.06\\
&I&1.70&	5.55&	-0.16&	0.48&	0.86&	0.09&	0.06\\
&Cl&2.08&	8.68&	-0.25&	0.45&	0.74&	0.10& 0.16\\
Sn&Br&1.73&	7.11&	-0.21&	0.50&	0.77&	0.11&	0.16\\
&I&1.68&	6.23&	-0.15&	0.48&	0.83&	0.10&	0.14\\
&Cl&2.46&	7.57&	-0.31&	0.49&	0.72&	0.10&	0.52\\
Pb&Br&2.36&	6.88&	-0.30&	0.52&	0.79&	0.11&	0.52\\
&I&2.18&	6.05&	-0.22&	0.48&	0.85&	0.10&	0.50\\
\hline
\hline
\end{tabular}
\label{T1}
\end{table}
\section{Spin Texture from the Toy Hamiltonian}
\begin{figure}
\centering
\includegraphics[angle=-0.0,origin=-1.0,height=12.0cm,width=7.0cm]{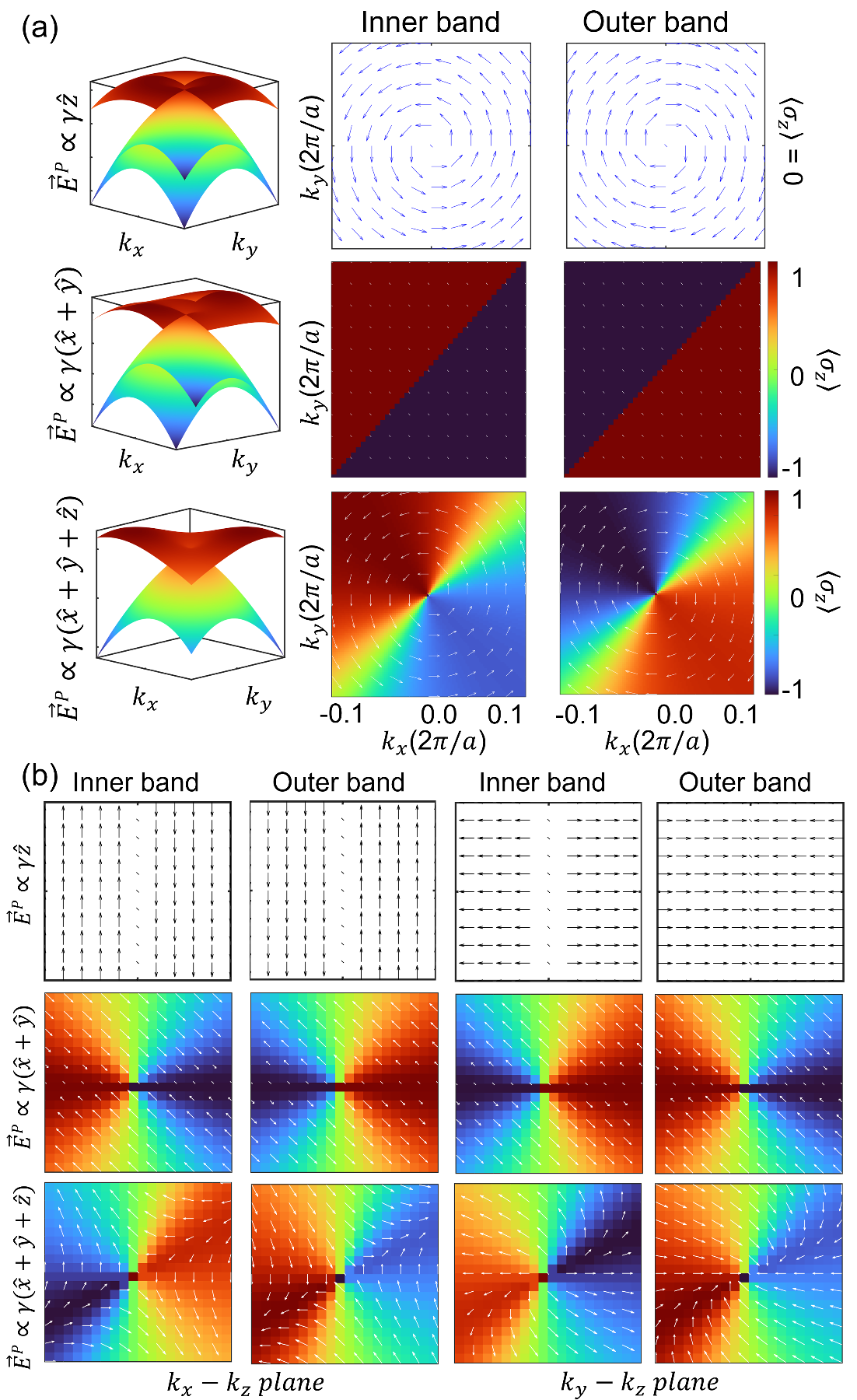}
\caption{(a; left) The 3D (inner and outer) valence bands for different polarised field. (a; right) The corresponding spin textures in the xy - plane. (b) The spin textures in the yz and xz planes for the considered polarized field. }
\label{figS1}
\end{figure}

The spin texture of the toy \textbf{k$\cdot$p} Hamiltonian (Eq. 7-11 in main text) are shown in Fig. \ref{figS1}. They reasonably reproduces the spin texture that of the downfolded Hamiltonian (Fig. \ref{figS2} in supplementary information (SI) and Fig. 5 in main text).
\section{Spin-Textures of Inner Valance Band at the Different Plane of Brillouin Zone}
\begin{figure}
\centering
\includegraphics[angle=-0.0,origin=-1.0,height=8.0cm,width=17cm]{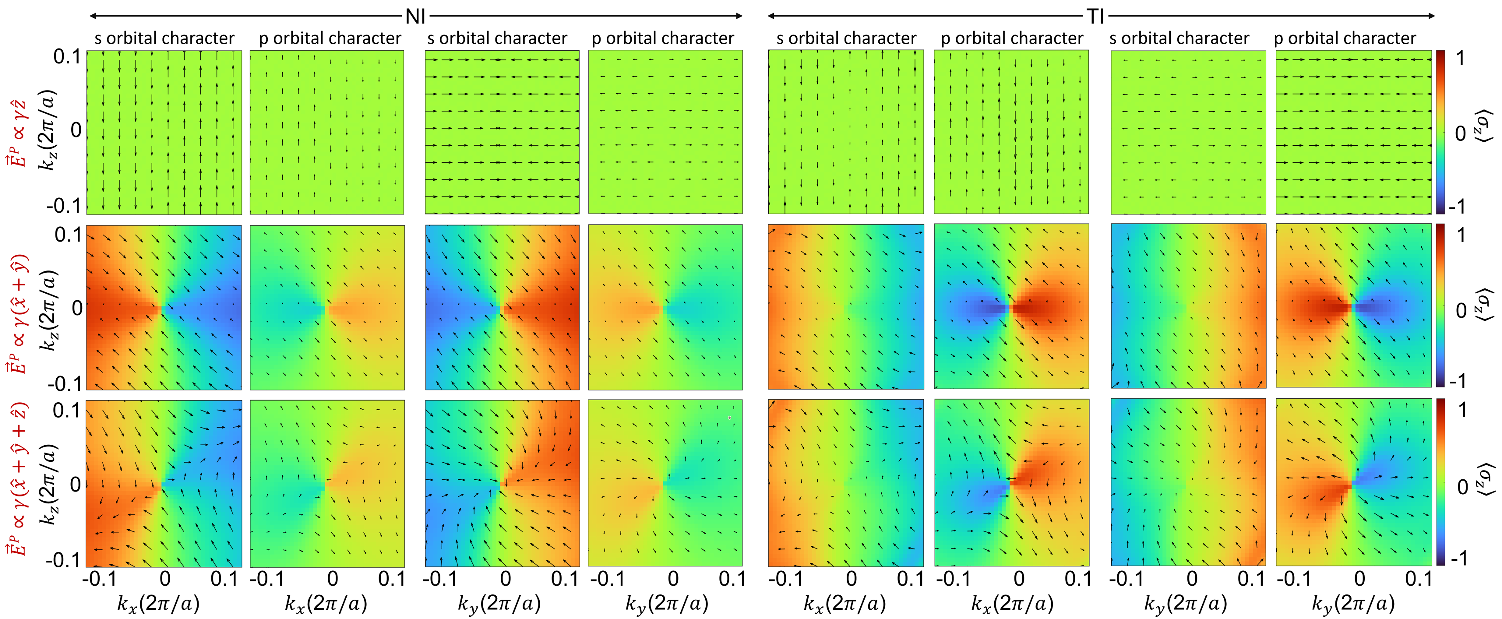}
\caption{The spin texture of inner valance band around the R-point in $k_x-k_z$ and $k_y-k_z$ plane for various polarization field direction. Field along $\hat{z}$ direction (first row), (Second row) Field in $(\hat{x}+\hat{y})$ direction and (third row) field in $(\hat{x}+\hat{y}+\hat{z})$ direction. The z-component of spin expectation ${\sigma_z}$ is shown by the background color gradient.}
\label{figS2}
\end{figure}

In Fig. \ref{figS2}, we have shown the spin texture under normal insulator (NI) and topological insulator (TI) phase for the inner valance band at the plane $k_x$-$k_z$ at $k_y = \pi$ and $k_y$-$k_z$ at $k_x = \pi$.
\begin{figure}
\centering
\includegraphics[angle=-0.0,origin=-1.0,height=7.0cm,width=17.0cm]{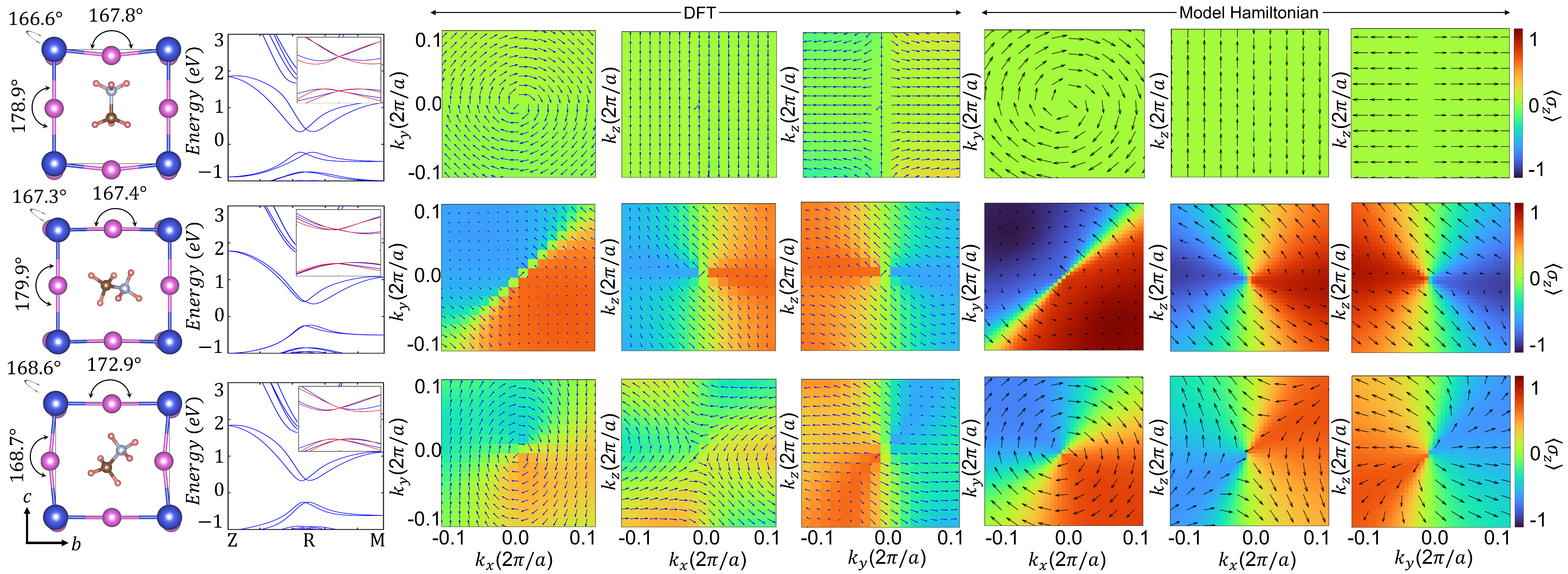}
\caption{ (First column from left) Relaxed cubic structure of MAPbI$_3$ for orientation of MA along [001], [110], and [111]. (Second column) Corresponding to these relaxed cells, DFT and model Hamiltonian (red; inset) band structures. (third to fifth columns)  Spin textures as obtained from DFT. (sixth to eighth columns) Spin textures as obtained from model Hamiltonian showing an excellent agreement with the DFT.}
\label{figS3}
\end{figure}
\section{DFT Study of Spin Textures and validation of the Model Hamiltonian}
As discussed in the main text, to further validate the model Hamiltonian, here we will compare the spin texture obtained from it with that of DFT. For this purpose, relaxed cubic MAPbI$_3$ (Pm3m, a = 6.313\AA) is considered for orientation of MA along [001], [110], and [111]. The relaxed cells, band structures, and spin textures are shown in Fig. \ref{figS3}. To bring the comparison, one needs to feed the polarization field as obtained from DFT to the model Hamiltonian. For this purpose, first, the ferroelectric polarization was calculated using the Berry phase method \cite{Pola1993}. Second, the ratio of the component of the polarization along $\hat{x}$, $\hat{y}$, and $\hat{z}$ is same as the component of the electric field which enters the Hamiltonian, Eq. 7, through $\{$ $\gamma_x$, $\gamma_y$, and $\gamma_z$ $\}$ (also see discussion in the main text). Therefore, the spin textures from the model Hamiltonian are obtained by using the optimized $\gamma$ values for all the three MA orientations as listed in Table. \ref{T2}. The DFT and model Hamiltonian spin textures are shown in Fig. \ref{figS3} and they have excellent agreement.
\begin{table}
\centering
\caption{ DFT calculated polarization field components at three different axes for PbI relaxed cubic MAPbI$_3$.}
\begin{tabular}{cccccccc}
\hline
\bf MA molecule & \bf Polarization &  ($ \bf \mu C/cm^2$)& & \bf Total Polarization &&&\\
\bf orientation & \bf $\hat{x}$ & \bf $\hat{y}$ & \bf $\hat{z}$ & ($ \bf \mu C/cm^2$) &$\gamma_x$&$\gamma_y$&$\gamma_z$\\
\hline
$[001]$ & 0.00 & 0.00 & -8.0164& 8.0164&0.0&0.0&-0.1\\
$[110]$ & 11.97 & 11.97& -3.502& 16.562&0.0723&0.0723&-0.0211\\
$[111]$ & 7.769 & 7.704 & 12.098&16.9742&0.0476&0.0472&0.0741\\
\hline
\end{tabular}
\label{T2}
\end{table}


\end{document}